\documentclass[pdftex]{article}
\usepackage{aas_macros}
\usepackage{icrctc07}

\title{A Fresh Look at Diffuse Gamma-ray Emission from the Inner Galaxy}
\shorttitle{Diffuse Gamma-ray Emission}
\authors{B. M. Baughman, W. B. Atwood, R. P. Johnson, T. A. Porter, and M. Ziegler}
\shortauthors{B. M. Baughman and et al}
\afiliations{Santa Cruz Institute for Particle Physics, University of California, Santa Cruz, CA 95064, U.S.A.\\ }
\email{brian@scipp.ucsc.edu}

\abstract{The Energetic Gamma-Ray Experiment Telescope (EGRET) experiment 
onboard the Compton Gamma-Ray Observatory (CGRO) 
has provided the most precise measurements of the \gray{} sky to date. 
The EGRET measurements of the diffuse emission across the sky show 
an excess above 1 GeV. 
This ``GeV excess'' has been a topic of great debate and interest since 
its original discovery by Hunter et al. in 1997. 
We have modified the GLAST simulation and reconstruction software to 
model the EGRET instrument. 
This detailed modeling has allowed us to explore the parameters of the 
EGRET instrument, 
in both its beam-test configuration and in-orbit on CGRO, 
in greater detail than has previously been published. 
We have found that the GeV excess is significantly increased when 
previously unaccounted for instrumental effects are considered. 
We will present a new measurement of diffuse \gray{} emission in the 
inner Galaxy.}

\newcommand{\degr}{\ensuremath{^\circ } }
\newcommand{\aeff}{\ensuremath{A_{eff} } }
\newcommand{\gray}[1]{$\gamma$-ray{#1}}

\begin{document}
\maketitle

\section{Introduction}
The EGRET\cite{1988SSRv...49...69K} 
telescope onboard the CGRO provided the most 
detailed look at the \gray{} sky to date. 
Before launch, the EGRET collaboration performed extensive 
testing \cite{1993ApJS...86..629T} to characterize the instrument response to 
both charged particles and \gray{s}. 
The instrument response functions found during these tests were the basis for 
constructing the EGRET exposure maps and thus the EGRET intensity sky maps. 
While there are references to a Monte Carlo simulation of the EGRET in a 
variety of EGRET publications \cite{1993ApJS...86..629T,1998ApJ...494..523S}, 
it is clear from those publications that the simulations were preliminary in 
nature.

The simulation environment developed for use with the Gamma-ray Large Area 
Space Telescope (GLAST) \cite{1994NIMPA.342..302A} provides a basis on 
which we have developed our simulations. 
The simulation framework is based on GEANT4 
\cite{Agostinelli:2002hh,Allison:2006ve} which is well tested. 
We have constructed a geometric model of the EGRET instrument based on the 
most detailed information available. 
We have also included parametric response models for the EGRET sub-detectors. 
Both the geometric and response models are incorporated into the framework. 
Furthermore, we have integrated the Burst and Transient Source Experiment 
(BATSE) Mass Model of CGRO \cite{2003A&A...398..391S}, 
used to estimate backgrounds in the BATSE instrument, with our model of the 
EGRET instrument.  

Characterization of the EGRET instrument was extensive, consisting of a 
charged particle beam test and two \gray{} beam tests. 
However, it was not possible to test all configurations of the instrument, 
nor was it possible to examine the particle interactions outside of the 
detectors. 
The Monte Carlo framework we have constructed allows us to probe the 
instrument in detail and access information about the simulated events 
not available in the laboratory. 
Thus, we have been able to compare the differences between the EGRET in the 
beam test environments to the EGRET in its flight environment. 
Most notably, 
we have probed effects relating to the integration of the EGRET onto CGRO. 

The EGRET was 
a pair-conversion telescope. 
As such, it required a method for rejecting charged particles entering the 
detector that might otherwise be treated as \gray{s}. 
This was accomplished by the anti-coincidence system known as the A-dome. 
The A-dome was a monolithic scintillator which was read out by 24 
photo-multiplier tubes (PMTs) optically coupled to the lowest edge of the dome. 
During the EGRET beam test at the Stanford Linear Accelerator Center (SLAC) 
it was 
discovered that the EGRET effective area (\aeff) decreased at a rate that 
was faster than expected at high energies. 
The decrease was determined to be caused by 
``self-veto'' \cite{1993ApJS...86..629T}, where an otherwise 
acceptable \gray{} event is vetoed by the A-dome when a secondary particle 
associated with the electromagnetic shower of the \gray{} in the instrument 
triggers a veto signal. 
The secondary particles can be either charged particle or X-rays which 
Compton scatter within the A-dome scintillator. 

Our simulation environment has allowed us to explore the problem of 
self-veto with greater detail than was available during the calibration of 
the EGRET. 
Furthermore, we have simulated the instrument in the environment that the 
astrophysics data were taken, specifically within close proximity to the CGRO. 
Comparison between the beam test geometry and the flight geometry has 
produced some interesting results, primarily we find that the effect of 
self-veto is exacerbated by the EGRET being attached to the CGRO.

\section{Analysis Methods}
The simulations have a variety of parameters which are related to the 
EGRET instrument response. 
Since we are primarily concerned with the effect of self-veto on the 
\aeff we limit our discussion to the relevant parameters. 
Effective vetoing of charged particle events passing through the A-dome 
was dependent on the voltage settings of the readout PMTs. 
While in-flight, it was necessary to reduce the efficiency of vetoing 
charged particles during calibration of the calorimeter. 
Thus, the voltage settings on the readout PMTs were adjustable. 
To replicate this, 
we have implemented a threshold for energy measured at the PMTs for our 
simulated events. 
By measuring the \aeff for incident mono-energetic \gray{s} at the same 
energies measured during the EGRET beam tests we are able to adjust this 
threshold for an optimized fit to the EGRET reported \aeff. 
We have also allowed for an overall normalization factor between the 
EGRET track finding algorithm and the one employed in our simulations. 

\begin{figure}
\begin{center}
\includegraphics [width=0.48\textwidth]{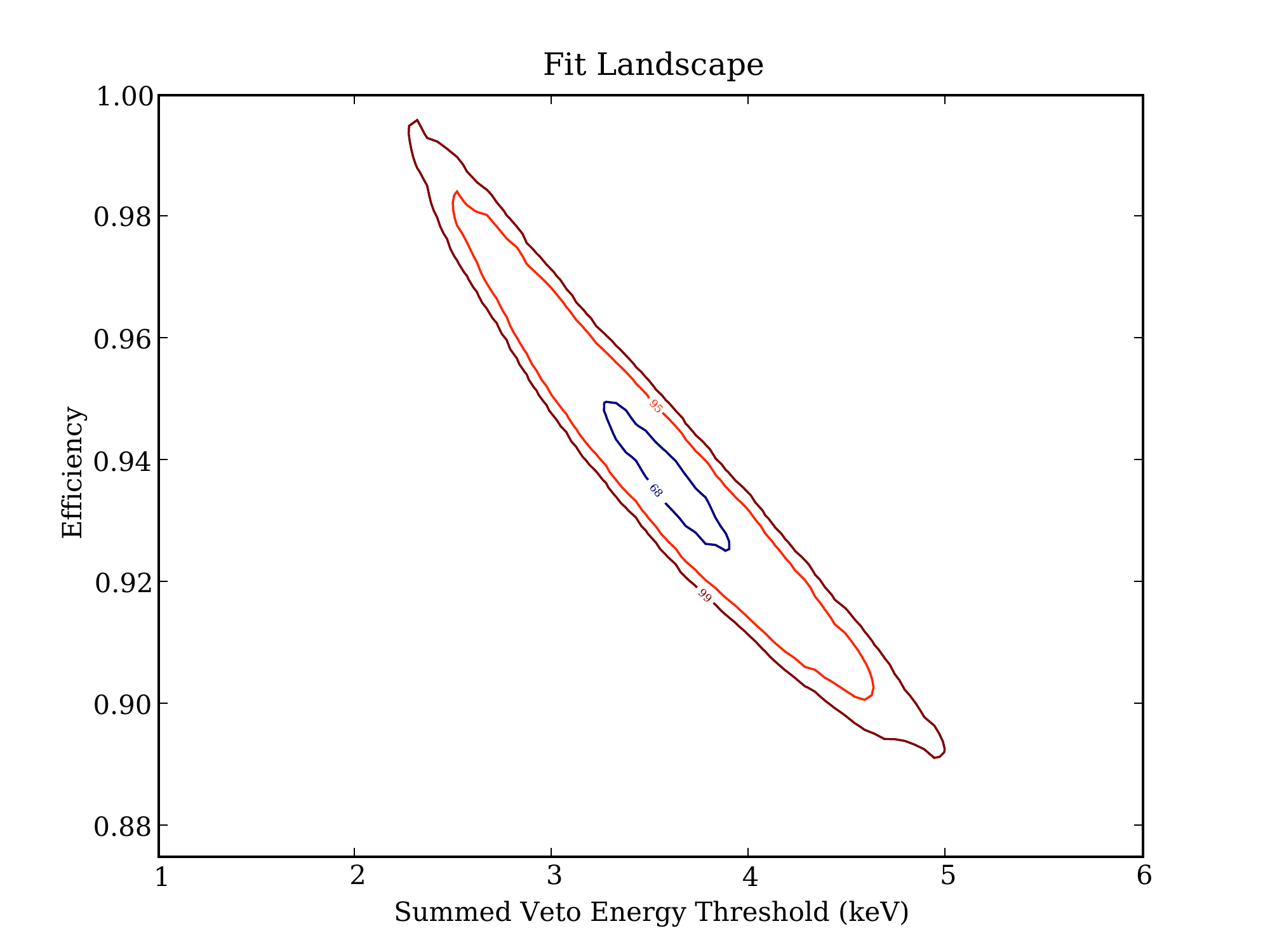}
\end{center}
\caption{Optimization landscape with 68\%, 95\%, and 99\% confidence 
contours. 
Summed veto threshold energy is calculated by summing the attenuated 
energy deposited within the A-dome.}
\label{fig:fitland}
\end{figure}

\begin{figure}
\begin{center}
\includegraphics [width=0.48\textwidth]{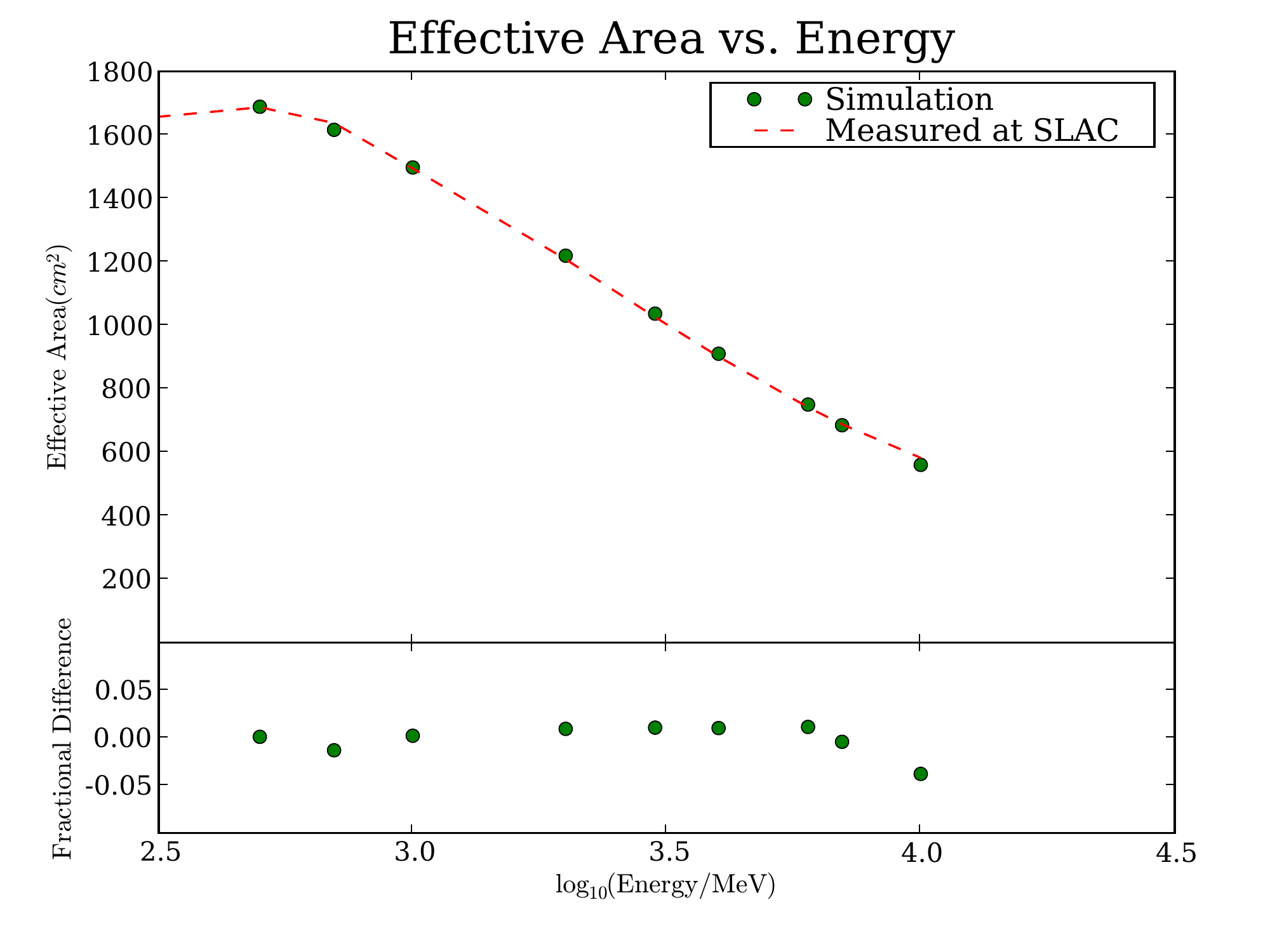}
\end{center}
\caption{Simulated \aeff after optimizing the summed veto threshold energy 
to the beam test results from SLAC compared to the measured values.}
\label{fig:fitaeff}
\end{figure}

The results of our optimization can be seen in Figs. \ref{fig:fitland} 
and \ref{fig:fitaeff}. 
This optimization was performed using simulated data in the beam geometry. 
Every attempt was made to mimic the EGRET triggering algorithm and 
spark chamber response. 
However, as was noted by the EGRET collaboration \cite{1993ApJS...86..629T}, 
below $300$ MeV there are noticeable effects on the track finding due to the 
difference in efficiencies between tracking layers, 
thus we have performed our fit only for energies $\ge 300$ MeV.

For both the beam and flight geometries we generated 800000 events at near 
normal incidence in each of the 10 standard EGRET energy bins,
plus three extended bins (10--20 GeV, 20--50 GeV, 50--120 GeV), 
using a spectrum proportional to $E^{-2.1}$; this corresponds to the 
weighting used for generation of the EGRET exposure maps. 

We construct scale factors for each of the 10 standard exposure energy 
ranges using the following formula:

\begin{equation}
\label{eqn:scalefactor}
F_i = \left(\frac{N^{accepted}_{i,flight}}{N^{generated}_{i,flight}}\right)\bigg/\left(\frac{N^{accepted}_{i,beam}}{N^{generated}_{i,beam}}\right)
\end{equation}

\noindent
where $N_i$ is the number of events in energy bin $i$, 
the subscript labels refer to the geometrical
configuration of the simulation (beam test or flight), 
and the superscript labels refer 
to the total number of generated events (generated) and the events accepted
for reconstruction (accepted). 
These factors are constructed assuming the same angular dependence as the 
published EGRET \aeff{}.


Extension of the EGRET effective area beyond 10 GeV has been done previously 
using the preliminary Monte Carlo mentioned earlier \cite{2005ApJS..157..324T}. 
Our extension uses a similar approach.
The exposure map for the 4--10 GeV bin is used as a basis for all 
higher energy bins.
The exposure maps for the energy bins 10--20 GeV, 20--50 GeV, and 50--120 GeV, 
are generated using a scaling factor for each higher energy bin 
that accounts for the relative difference 
between the number of events generated and accepted for reconstruction 
in that bin compared
with the number of events for the 4--10 GeV bin. 
The following formula is used to create the scaling factors:

\begin{eqnarray}
\label{eqn:hescalefactor}
F_{HE,j} &=& \left(\frac{N^{accepted}_{j,flight}}{N^{generated}_{j,flight}}\right)/A_{9} \\
A_{9} &=& \left(\frac{N^{accepted}_{9,beam}}{N^{generated}_{9,beam}}\right) 
\end{eqnarray}

\noindent
where $j =$ 10--20, 20--50, 50--100 GeV, the superscripts and 
subscripts have the same meaning as in Eq.~\ref{eqn:scalefactor}, and 
$N^{accepted} _{9,beam}$ and $N^{generated} _{9,beam}$ are the number 
of accepted and generated events in the beam configuration for the 4--10 GeV 
bin, respectively.

Constructing the scaling factors as described reduces the possibility 
of systematic effects unrelated to changes in geometry. 
The ratio of flight geometry performance to beam test geometry performance 
should be invariant to effects unrelated to the change in geometry. 
The performance for each geometry has been examined in detail and there are 
no unanticipated effects that might introduce large systematic errors to our 
results.

\section{Results}
The scale factors calculated above are applied to the corresponding 
EGRET exposure maps. 
All the scaling factors found imply a systematically lower \aeff at high 
energies. 
This implies that each \gray{} measured is more significant than previously 
thought, leading to a systematic increase in both the integrated flux as 
well as the hardness of the spectra measured by the EGRET. 

\begin{figure}
\begin{center}
\includegraphics [width=0.48\textwidth]{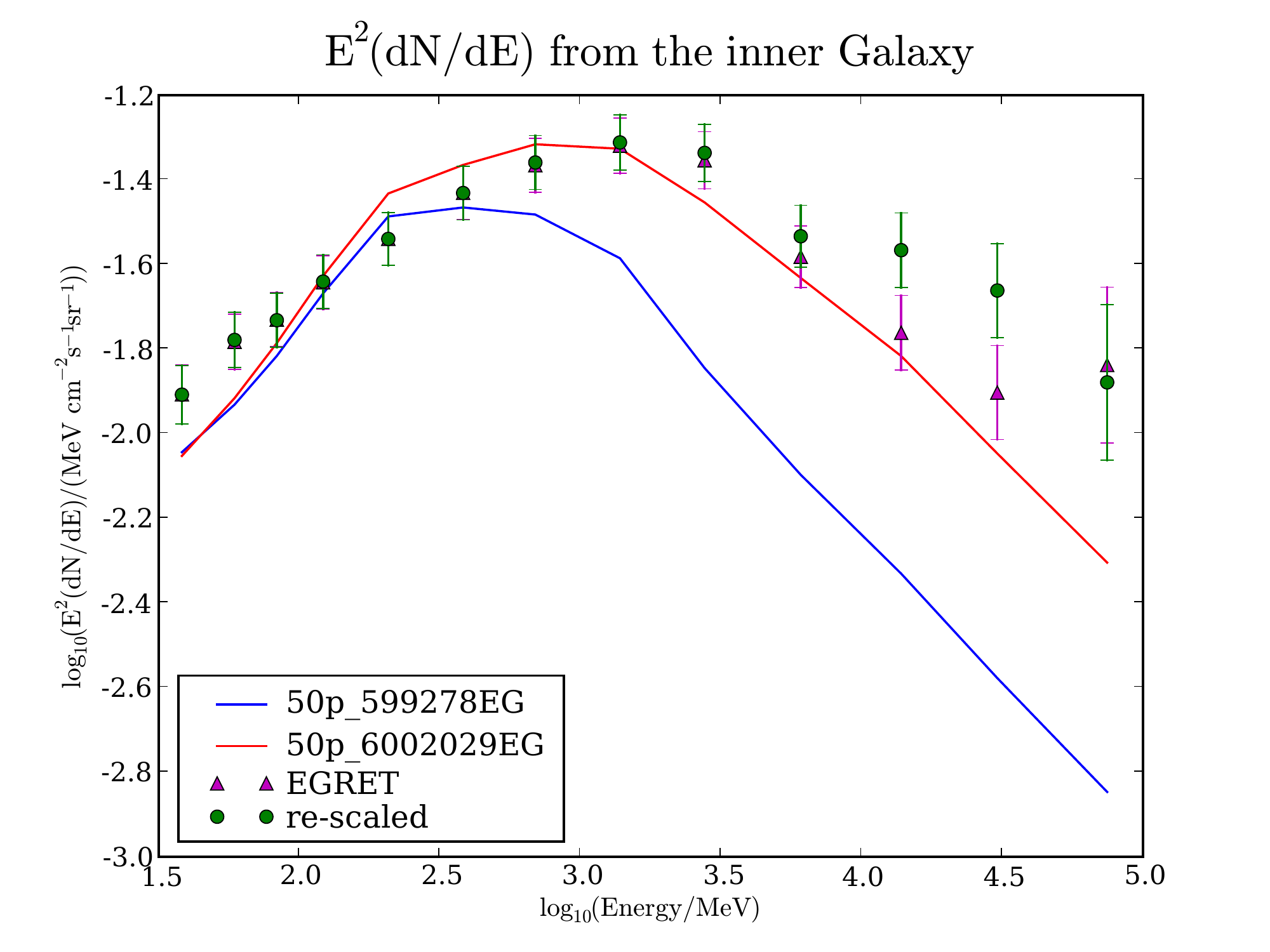}
\end{center}
\caption{Comparison of $E_\gamma^{2} dN_\gamma/dE_\gamma$ spectra for 
$0.5\degr<|l|<30\degr$ and $0.5\degr<|b|<6\degr$.}
\label{fig:igesqspec}
\end{figure}

We have re-analyzed the EGRET data set in the inner 
Galaxy, $0.5\degr<|l|<30\degr$ and $0.5\degr<|b|<6\degr$, using our re-scaled 
exposure maps. 
For comparison we have also analyzed two commonly accepted 
GALPROP \cite{1998ApJ...509..212S} models, 
599278 \cite{2004ApJ...613..962S} and 
6002029RE \cite{2007ApJ...658L..33A}. 
The GALPROP model 599278 is constructed by assuming the observed radiation, 
gas, and cosmic-ray distributions are representative of the Galaxy at 
large. 
This is referred to as the ``conventional'' model.  
6002029RE has been modified to better reproduce the EGRET observations, 
specifically ``secondary antiproton data were used to fix the Galactic 
average proton spectrum, while the electron spectrum is adjusted using the 
spectrum of diffuse emission itself''\cite{2004ApJ...613..962S}. 
This is known as the ``optimized'' model. 
These models were run in their published configurations, except we changed the
energy and skymap binning to correspond to the EGRET skymaps.
In the following, we denote the runs for these models as 
599278EG and 6002029EG, respectively.
Fig.~\ref{fig:igesqspec} shows the $E_\gamma^{2} dN_\gamma/dE_\gamma$ 
for our 599278EG and 
6002029EG GALPROP models as well as the re-scaled and original EGRET 
measurements. 

\section{Conclusions}
The re-scaled EGRET spectrum for the inner Galaxy shows an increased excess
when compared to previous results.
When compared with the conventional model, the 
EGRET excess has a reduced $\chi^{2}$ of 18.8 and 24.7 
for the original and re-scaled EGRET measurements, respectively.
For the optimized model, the
EGRET excess has a reduced $\chi^{2}$ of 1.9 and 3.4 for the 
original and re-scaled EGRET measurements, respectively. 
This analysis indicates that the GeV excess may be larger than 
previously thought. 
It is important to note that while the GeV excess is dramatic with respect to 
the conventional GALPROP model (599278EG) it is much reduced with respect to 
the optimized model (6002029EG).  

\section{Acknowledgments}
We would like to thank the thoughtful criticism and suggestions of 
Gottfried Kanbach (MPE), Seth Digel (SLAC), Olaf Reimer (Stanford), 
Andy Strong (MPE), Donald Kniffen (GSFC), and Dave Thompson (GSFC), 
all of whose input proved invaluable.

This work is partially supported by the United States Department of 
Energy through grant DE-FG02-04ER41286. 
Further support was provided by NASA grant PY-1775.

\bibliography{./References}

\bibliographystyle{h-physrev}

\end{document}